%% file: HorizonsDraft.tex
\documentclass[12pt]{article}
\pdfoutput=1 

\usepackage{jheppub} 

\usepackage[T1]{fontenc} 
\usepackage{amsmath,physics}
\usepackage{comment}
\usepackage{bbm}
\usepackage[percent]{overpic}
\usepackage[toc,page]{appendix}
\usepackage{etoolbox}
\usepackage[final]{showlabels}
\usepackage{tabularx}
\usepackage{bookmark}

\BeforeBeginEnvironment{appendices}{\clearpage}
\AfterEndEnvironment{appendices}{\clearpage}
\AfterEndEnvironment{appendices}{\pretocmd{\section}{\clearpage}{}{}}{}
\AfterEndEnvironment{appendices}{\pretocmd{\subsubsection}{\clearpage}{}{}}{}

\usepackage{transparent} 
\usepackage{mathrsfs} 
\usepackage{mathtools} 

\usepackage{float,caption,subcaption,xcolor}
\usepackage{graphicx}
\usepackage{hyperref}
\hypersetup{
    colorlinks=true,
    linkcolor=blue,
    filecolor=magenta,      
    urlcolor=cyan,
    pdftitle={Black Hole Horizons from the Double Copy},
    }
\usepackage{multirow}

\usepackage{bbm}

\title{Black Hole Horizons from the Double Copy}
\author[a]{Samarth Chawla}
\author[a]{and Cynthia Keeler}

\affiliation[a]{Department of Physics, Arizona State University,
Tempe, AZ 85281, USA}

\emailAdd{samarthc@asu.edu}
\emailAdd{keelerc@asu.edu}

\abstract{We describe a procedure for locating black hole horizons in Kerr-Schild spacetimes in the double copy paradigm. Using only single- and zeroth-copy data on flat spacetime, our procedure predicts the existence of trapped surfaces in the double-copy gravitational solution. We show explicitly how this procedure locates the horizon of the Schwarzschild black hole and the general Myers-Perry black hole.}

\begin{document} 
\maketitle
\flushbottom

\section{Introduction} \label{sec:intro} 

The classical double copy, in its most straightforward formulation \cite{Monteiro:2014cda}, maps classical
solutions of the Abelian sector of Yang-Mills theory and classical solutions of a biadjoint scalar
theory to classical solutions of general relativity. The gravitational solution is expressed as two copies of the
gauge theory solution, hence the moniker ``double copy''. Conversely, the gauge solution is often called the
``single copy'' of the gravity solution, while the scalar is termed the ``zeroth copy''. The underpinnings of this
double copy procedure lie in the color-kinematics duality between gauge and gravity amplitudes (see
\cite{Travaglini:2022uwo,Bern:2019prr,Adamo:2022dcm} for recent reviews). Since the original formulation of the
double copy for Kerr-Schild spacetimes \cite{Monteiro:2014cda}, several other examples of classical double copy
relations have been found
\cite{Luna:2015paa,Luna:2017dtq,Ridgway:2015fdl,Adamo:2017nia,Bahjat-Abbas:2017htu,CarrilloGonzalez:2018ejf,Ilderton:2018lsf,Gurses:2018ckx,Carrillo-Gonzalez:2017iyj,Lee:2018gxc,Lescano:2020nve,Lescano:2021ooe,Bah:2019sda,Goldberger:2019xef,Kim:2019jwm,Luna:2020adi,Easson:2020esh,Berman:2018hwd,Alkac:2021seh,Mkrtchyan:2022ulc,Adamo:2022rob,Alfonsi:2020lub,Adamo:2020qru,Gonzo:2021drq,Luna:2018dpt,Keeler:2020rcv,Easson:2021asd,Easson:2022zoh,Godazgar:2020zbv,White:2020sfn,Chacon:2021wbr,Chacon:2021lox,Han:2022mze,Alawadhi:2020jrv,Godazgar:2021iae,Armstrong-Williams:2023ssz}.
For recent reviews of the classical double copy, see \cite{Kosower:2022yvp} and chapter 8 of \cite{Bern:2019prr}.

The simplest non-trivial example of the classical double copy is a map between the Schwarzschild metric, which can
be interpreted as the gravitational field of a static point mass, and
the gauge field of a static point charge. While it is unsurprising that these static and spherically symmetric
solutions map to each other, the distinctive event horizon of the Schwarzschild
metric seems to have no analogue in its single copy, the Coulomb solution. One of the broader aims of the entire
double copy program is to describe quantities in gravity and gauge theory using a common set of building blocks
\cite{Bern:2019prr}. Thus a natural question to ask is: how does the Schwarzschild event horizon emerge out of its
double copy construction from the Coulomb potential?

The teleological definition of a black hole (see for instance \cite{Hawking:1973uf}) is the
region of spacetime outside the causal past of future null infinity. The boundary of such a region, if it exists,
is necessarily a null hypersurface and is termed the event horizon. While this definition effectively formalizes
the notion of a black hole as a region of no escape, it is a highly non-local definition that relies on knowledge
of the global structure of the metric all the way to asymptotic infinity. This non-locality presents a potential
obstacle in finding event horizons from the double copy, which is a local map.\footnote{While the classical double copy in the formulations we have relied on is indeed local in position space, the double copy of amplitudes is local in momentum space. This raises the question of why the classical version is a local product rather than a convolution, a question addressed recently in \cite{Monteiro:2021ztt, Luna:2022dxo}.} The statement of the Kerr-Schild
double copy \eqref{kerrSchildMetricFromA}, for instance, maps the gauge field $A_{\mu}$ at each point to the metric
$g_{\mu \nu}$ at the same point.

Fortunately, several local definitions of horizons have been proposed in the literature. See
\cite{Gourgoulhon:2008pu,Ashtekar:2004cn,Booth:2005qc,Senovilla:2011fk} for reviews of various local horizon
definitions. All local horizons found therein have one feature in common: the horizon hypersurface must possess a
foliation in terms of marginally trapped surfaces (we review the definition of trapped surfaces in section
\ref{sec:localHorizonDefs}). Thus, to remain agnostic towards the various local horizon definitions, our general
discussion focuses primarily on deducing the existence of trapped surfaces in the double copy spacetime using only
single- and zeroth-copy data. 

In the following section, we review the Kerr-Schild formulation of the classical double copy and briefly survey a
few local horizon definitions. Section \ref{sec:setup} builds up to a derivation of our main result: a prescription
for finding local horizons using only single- and zeroth-copy calculations on an auxiliary flat spacetime. We then
illustrate the use of the general prescription by constructing the event horizons of the well-studied Schwarzschild
and Myers-Perry solutions in section \ref{sec:examples}. Finally, in section \ref{sec:discussion} we discuss our
result and directions for future work.

\section{Review} \label{sec:review} 

\subsection{Classical Double Copy} \label{sec:classicalDoubleCopy}
We review in this section the classical double copy for Kerr-Schild spacetimes, first studied in
\cite{Monteiro:2014cda}. A Kerr-Schild spacetime is one whose metric can be written as
\begin{equation} \label{kerrSchildMetric} 
    g_{\mu \nu} = \eta_{\mu \nu} + \Phi k_{\mu} k_{\nu},
\end{equation}
where $\eta_{\mu \nu}$ is a metric with zero curvature, and $k_{\mu}$ is null and geodesic. The inverse metric is
\footnote{We note here that the index on $\eta_{\mu \nu}$ is not raised using the metric, that is $\eta^{\mu \nu}
\neq g^{\mu \rho} g^{\nu \lambda} \eta_{\rho \lambda}$. Instead, it is defined to satisfy $\eta^{\mu \rho}
\eta_{\rho \nu} = \delta^{\mu}_{\nu}$. }
\begin{equation} \label{kerrSchildMetricInverse} 
    g^{\mu \nu} = \eta^{\mu \nu} - \Phi k^{\mu} k^{\nu}.
\end{equation}
The index on $k_{\mu}$ can be raised with
either $g_{\mu \nu}$ or $\eta_{\mu \nu}$, since
\begin{equation} 
    g^{\mu \nu} k_{\mu} = \eta^{\mu \nu} k_{\mu} - \Phi \left(k^{\mu} k_{\mu}\right) k^{\nu} =
    \eta^{\mu \nu} k_{\mu}.
\end{equation}
Thus, we can unambiguously write
\begin{equation} \label{kNull} 
    k^{\mu} k_{\mu} = 0.
\end{equation}
Similarly, $k_{\mu}$ is geodesic with respect to both the flat metric $\eta_{\mu \nu}$ or the curved metric $g_{\mu
\nu}$,
\begin{equation} \label{kGeodesic} 
    k^{\mu} \nabla_{\mu} k_{\nu} = k^{\mu} \mathring{\nabla}_{\mu} k_{\nu} = 0,
\end{equation}
where $ \mathring{\nabla}_{\mu}$ denotes the Levi-Civita connection corresponding to the metric $\eta_{\mu \nu}$.

The single copy field is the Abelian potential $A_{\mu} = \Phi k_{\mu}$, and $\Phi$ is the zeroth copy scalar
field. They satisfy Maxwell's equations and the massless Klein-Gordon equation respectively on the flat background
$\eta_{\mu \nu}$ whenever the metric $g_{\mu \nu}$ satisfies the vacuum Einstein equations.\footnote{
\label{FlatVCurvedFootnote}  Much of the double copy literature uses a flat background owing to its origins in the
study of amplitudes in flat spacetime. However, the classical double copy can still be performed in curved
spacetimes, and the curved equations of motion can be used; for relevant literature, see \cite{Alawadhi:2020jrv,
Gurses:2018ckx,Bahjat-Abbas:2017htu, Carrillo-Gonzalez:2017iyj, Prabhu:2020avf, Adamo:2017nia, Alkac:2021bav,
Han:2022mze,Han:2022ubu}.}

The classical double copy construction of Kerr-Schild spacetimes thus maps certain classical solutions of the
Maxwell and Klein-Gordon equations to classical solutions of the vacuum Einstein equations. The metric is 
constructed ``on top of'' the flat metric $\eta_{\mu \nu}$ using a combination of $A_{\mu}$ and $\Phi$,
\begin{equation} \label{kerrSchildMetricFromA} 
    g_{\mu \nu} = \eta_{\mu \nu} + \frac{1}{\Phi} A_{\mu} A_{\nu}.
\end{equation}
There are other formulations of the double copy, most notably the Weyl double copy introduced in
\cite{Luna:2018dpt}. However, we limit ourselves in scope to Kerr-Schild spacetimes and the Kerr-Schild double copy.

As pointed out in the introduction, the teleological definition of a black hole horizon requires knowledge of the
global structure of spacetime, while the Kerr-Schild double copy is a local map. To be able to find horizons using
the double copy we thus need a local definition of black hole horizons, which we briefly review in the next
section.

\subsection{Local definitions of black hole horizons} \label{sec:localHorizonDefs}

Various local definitions of black hole horizons have been advanced in the literature. 
In this section we review relevant material from \cite{Gourgoulhon:2008pu}. All local horizon definitions therein
rely on the notion of a trapped surface. Intuitively, a co-dimension $2$ spacelike surface is a trapped surface
when all light rays leaving the surface are condemned by the ambient geometry to always fall ``inward''. The notion
of always falling ``inward'' is made precise by looking at how the co-dimension $2$ volume element evolves along
null congruences leaving the surface.

Any co-dimension $2$ spacelike surface $\mathcal{S}$  has, at each point, two distinct future-directed null vectors
(call them $n^{\mu}$ and $l^{\mu}$) that are normal to it. Using the conventional normalization $n_{\mu} l^{\mu} =
-1$, we can write the induced metric on $\mathcal{S}$ as
\begin{equation}
    h_{\mu \nu} = g_{\mu \nu} + n_\mu l_\nu + l_\mu n_\nu.
\end{equation}
The co-dimension $2$ volume element is then $\sqrt{\det h}$. The expansion $\theta^{(V)}$ of $\mathcal{S}$ along a normal vector $V^\mu$ is defined as
\begin{equation} \label{expansionDefn}
    \theta^{(V)} = \mathcal{L}_V \ln \sqrt{\det h} = h^{\mu \nu} \nabla_\mu V_\nu,
\end{equation}
where $\mathcal{L}_V$ denotes the Lie derivative along $V^\mu$.

The surface $\mathcal{S}$ is a future-trapped surface whenever both $\theta^{(n)}$ and $\theta^{(l)}$ are negative.
Physically, this means that the cross sectional volume of a bundle of null trajectories leaving $\mathcal{S}$
always decreases, even along the nominally outgoing congruence $l^\mu$. The limiting case, where $\theta^{(l)}$
vanishes, is a marginally trapped surface.  In flat spacetime, the expansions along $n^\mu$ and $l^\mu$ have
opposite sign. The infalling congruence is the one with negative expansion, and the outgoing congruence has
positive expansion. We will assume throughout that $l^\mu$ is outgoing. 

A marginally trapped surface is thus one for which the cross section of the outgoing congruence $l^{\mu}$ is
instantaneously constant. Say we have a hypersurface $\mathcal{H}$ foliated by marginally trapped surfaces. Such a
surface is in general a trapping horizon. If
additionally the outermost condition $\mathcal{L}_{n} \theta^{(l)} < 0$ is satisfied, $\mathcal{H}$ is a
future outer trapping horizon (FOTH) \cite{Hayward:1993wb}. The outermost condition ensures that the expansion
$\theta^{(l)}$ turns negative inside $\mathcal{H}$, which means that $\mathcal{H}$ encloses a trapped region of
spacetime. If the outermost condition is not satisfied, $\mathcal{H}$ could still be a local horizon in terms of
the other definitions outlined in \cite{Gourgoulhon:2008pu}. For simplicity, in this paper we pick examples where
the notion of FOTH suffices.

\section{Finding horizons in a Kerr-Schild spacetime} \label{sec:setup}

In this section we describe, for a Kerr-Schild spacetime, how to search for the existence of horizons using
quantities calculated on a flat spacetime. We do so by foliating a given hypersurface into co-dimension $2$
surfaces $\mathcal{S}_{\lambda}$ using the Kerr-Schild vector $k_{\mu}$ \eqref{kerrSchildMetric} and computing
their null expansions, again using only flat spacetime quantities. This auxiliary flat spacetime is precisely the
one that appears in \eqref{kerrSchildMetric} and serves as the background spacetime for the single and zeroth copy
theories.

\subsection{The geometry of hypersurface foliations}
\begin{figure}[h]
    \centering
    \begin{subfigure}[t]{0.55\textwidth}
        \centering
        \def\svgwidth{\columnwidth}
        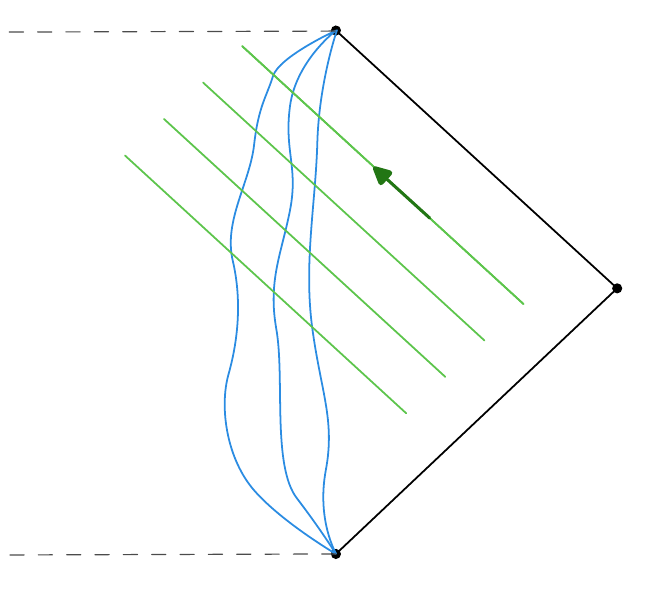
        \caption{Penrose diagram for an asymptotically flat spacetime, with a one-parameter family of hypersurfaces
        $\mathcal{H}_{(\xi)}$. Trajectories in green are integral curves of $k^{\mu}$. $\mathcal{S}_{1}$ and
    $\mathcal{S}_{2}$ are leaves of the foliation normal to $k^{\mu}$ on $\mathcal{H}_{1}$ and $\mathcal{H}_{2}$.}
    \end{subfigure}\hspace{2em}%
    \begin{subfigure}[t]{0.32\textwidth}
        \centering
        \def\svgwidth{\columnwidth}
        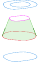
        \caption{Diagram of $\mathcal{H}_{1}$ and $\mathcal{H}_{2}$ with one angular direction restored.
        Schematically shows the evolution from $\mathcal{S}_{1}$ to $\mathcal{S}_{2}$ along $k^{\mu}$.}
    \end{subfigure}
    \caption{Foliation of hypersurfaces using $k_{\mu}$.}
    \label{figGeneric}
\end{figure}

In flat spacetime, let us pick an arbitrary hypersurface $\mathcal{H}$ with normal one-form $\tilde{H}_{\mu}$.
We will foliate $\mathcal{H}$ into spacelike surfaces $\mathcal{S}_{\lambda}$ parametrized by $\lambda$, with each
leaf of the foliation of co-dimension $2$ with respect to the background spacetime. At every point on a
co-dimension $2$ spacelike surface there are two independent future-directed null normals. We pick the foliation of
$\mathcal{H}$ such that $k_{\mu}$ can always be chosen as one of the two null normals. We also assume that $k^\mu$
is infalling, so that in the convention established in section \ref{sec:localHorizonDefs}, we
set $n^\mu = k^\mu.$ See figure \ref{figGeneric} for an illustration of the foliation picked out by $k^{\mu}$.

The normal subspace of each $\mathcal{S}_{\lambda}$ is by definition spanned by $k_{\mu}$ and $\tilde{H}_{\mu}$,
assuming the two to be linearly independent. Thus we can find the second null normal $l^{(0)}_{\mu}$ uniquely by
looking for vectors in the span of $k_{\mu}$ and $\tilde{H}_{\mu}$ which satisfy
\begin{equation} \label{l0Conditions} 
    l^{(0)}_{\mu} l^{(0)}_{\nu} \eta^{\mu \nu} = 0, \qquad l^{(0)}_{\mu} k_{\nu} \eta^{\mu \nu} = -1.
\end{equation}
The flat metric $\eta^{\mu \nu}$ is used to contract indices here since we are constructing the null normal
$l^{(0)}_{\mu}$ in flat spacetime. Note that since the Kerr-Schild vector $k_{\mu}$ satisfies $g^{\mu \nu} k_{\mu}
= \eta^{\mu \nu} k_{\mu}$, in all that follows we raise and lower indices on $k$ with impunity. The two conditions \eqref{l0Conditions} are enough to specify $l^{(0)}_{\mu}$ uniquely:
\begin{equation} \label{} 
    l^{(0)}_{\mu} = \frac{\tilde{H}_{\rho} \tilde{H}_{\sigma} \eta^{\rho \sigma}}{2 (k^{\rho}
    \tilde{H}_{\rho})^{2}} k_{\mu} - \frac{\tilde{H}_{\mu}}{k^{\rho} \tilde{H}_{\rho}}.
\end{equation}
The case where $k^{\rho} \tilde{H}_{\rho} = 0$ has the integral curves of $k^{\mu}$ lying on the $\mathcal{H}$, and
there is no foliation of the hypersurface with spacelike surfaces normal to $k_{\mu}$. As long as $k^{\rho}
H_{\rho} \neq 0$ then, we define for simplicity of notation
\begin{equation} \label{HNorm} 
    H_{\mu} = -\frac{\tilde{H}_{\mu}}{k^{\rho} \tilde{H}_{\rho}},
\end{equation}
and use $H_{\mu}$ as the hypersurface normal. Rewriting $l^{(0)}_{\mu}$ in terms of $H_{\mu}$,
\begin{equation} \label{l0FromH} 
    l^{(0)}_{\mu} = \frac{H_{\rho}H_{\sigma}\eta^{\rho\sigma}}{2} k_{\mu} + H_{\mu}.
\end{equation}

The intrinsic geometry of each $\mathcal{S}_{\lambda}$, inherited from $\eta_{\mu \nu}$, is given by the induced metric and its inverse,
\begin{equation} \label{flatInducedMetric} 
    \begin{split}
        \mathring{h}_{\mu \nu} &= \eta_{\mu \nu} + k_{\mu} l^{(0)}_{\nu} + l^{(0)}_{\mu} k_{\nu}, \\
        \mathring{h}^{\mu \nu} &= \eta^{\mu \nu} + k^{\mu} l^{(0)}_{\rho} \eta^{\nu \rho} + l^{(0)}_{\rho}
        \eta^{\rho \mu} k^{\nu},
    \end{split}
\end{equation}
where we use a circular accent to denote quantities defined on flat spacetime.

We now need to map $\mathcal{H}$ and its chosen foliation to the Kerr-Schild spacetime with metric $g_{\mu \nu}$
\eqref{kerrSchildMetric}. Since $\mathcal{H}$ and its foliation $\mathcal{S}_{\lambda}$ are described in terms of
the differential forms $H_{\mu}$ and $k_{\mu}$, they are in fact independent of the metric. Thus we can use the
same hypersurface $\mathcal{H}$ and its foliation $\mathcal{S}_{\lambda}$ with metric $g_{\mu \nu}$.
However, with a different metric the null normals of $\mathcal{S}_{\lambda}$ may no longer be the same.

Since $k_{\mu}$ is null with respect to both $\eta_{\mu \nu}$ and $g_{\mu \nu}$, it is still one of the null
normals. As in the previous case, the second null normal $l_{\mu}$ can be found in the span of $k_{\mu}$ and $H_{\mu}$ by imposing
\begin{equation} \label{lConditions} 
    l_{\mu} l_{\nu} g^{\mu \nu} = 0, \qquad l_{\mu} k_{\nu} g^{\mu \nu} = -1.
\end{equation}
Using the relation \eqref{kerrSchildMetricInverse} between $g^{\mu \nu}$ and $\eta^{\mu \nu}$, these conditions
specify $l_{\mu}$ uniquely as
\begin{equation} \label{lFroml0Down} 
    l_{\mu} = l^{(0)}_{\mu} - \frac{\Phi}{2} k_{\mu}.
\end{equation}
The intrinsic geometry of $\mathcal{S}_{\lambda}$ inherited from the metric $g_{\mu \nu}$ is given by the
induced metric and its inverse\footnote{Interestingly, owing to the
definition of $l_{\mu}$ in terms of $l^{(0)}_{\mu}$ \eqref{lFroml0Down} and the form of the metric \eqref{kerrSchildMetric}, it turns
out that $h_{\mu \nu} = \mathring{h}_{\mu \nu}$.}
\begin{equation} \label{curvedInducedMetric} 
    \begin{split}
        h_{\mu \nu} &= g_{\mu \nu} + k_{\mu} l_{\nu} + l_{\mu} k_{\nu} \\
        h^{\mu \nu} &= g^{\mu \nu} + k^{\mu} l_{\rho} g^{\nu \rho} + l_{\rho} g^{\rho \mu} k^{\nu}.
    \end{split}
\end{equation}

\subsection{The dynamics of hypersurface foliations}
Having described the choice of foliation, we now examine how each leaf $\mathcal{S}_{\lambda}$ evolves along its 
null normals. We specifically look at the expansion, defined as in \eqref{expansionDefn}.
Using the induced metric from \eqref{flatInducedMetric} we can write the expansions as
\begin{equation} \label{flatExpansions} 
    \mathring{\theta}^{(k)} = \mathring{h}^{\mu \nu} \mathring{\nabla}_{\mu} k_{\nu}, \qquad
    \mathring{\theta}^{(l^{(0)})} = \mathring{h}^{\mu \nu} \mathring{ \nabla}_{\mu} l^{(0)}_{\nu}.
\end{equation}
In flat spacetime there are no trapped surfaces, and thus the two expansions are always of opposite sign. Since we
picked the convention that $k^{\mu}$ is ingoing and $l^{\mu}$ is outgoing, $\mathring{\theta}^{(k)} \leq 0$ and
$\mathring{\theta}^{(l^{0})} \geq 0$. In other words, $k_{\mu}$ describes a converging null congruence and
$l^{(0)}_{\mu}$ describes a diverging null congruence.

In curved spacetime with metric $g_{\mu \nu}$, we use the induced metric \eqref{curvedInducedMetric} to calculate
the expansions as
\begin{equation} \label{curvedExpansions} 
    \theta^{(k)} = h^{\mu \nu} \nabla_{\mu} k_{\nu}, \qquad \theta^{(l)} = h^{\mu \nu} \nabla_{\mu} l_{\nu}.
\end{equation}
Here, the two may have the same sign. We are interested in spacetimes where $\theta^{(k)} \leq 0$ everywhere and
$\theta^{(l)} \geq 0$ near asymptotic infinity. Regions where $\theta^{(l)} < 0$ are trapped, and the boundary of such
regions, where $\theta^{(l)} = 0$, are trapping horizons.

\subsection{The double copy relation} \label{sec:doubleCopyRelation}
An explicit computation of the expansions using the metric \eqref{kerrSchildMetric} (outlined in appendix
\ref{app:ExpansionCalculation}) shows that for Kerr-Schild spacetimes the expansions on the $\eta_{\mu \nu}$ and
the $g_{\mu \nu}$ background are related via
\begin{equation} \label{thetaFromTheta0} 
    \theta^{(k)} = \mathring{\theta}^{(k)}, \qquad \theta^{(l)} = \mathring{\theta}^{(l^{(0)})} + \frac{\Phi}{2}
    \mathring{\theta}^{(k)}.
\end{equation}
Our assertion that $\theta^{(k)} \leq 0$ everywhere is automatically satisfied. We also see that there is a trapped
region if the term $\frac{\Phi}{2} \mathring{\theta}^{(k)}$ becomes larger in magnitude than
$\mathring{\theta}^{(l^{(0)})}$.

These relations constitute a ``double copy'' map in the sense that they relate the $\eta_{\mu \nu}$ and $g_{\mu \nu}$
dynamics of the foliation of $\mathcal{H}_{(\xi)}$ picked out by the Kerr-Schild vector $k^{\mu}$. Note that this
choice of foliation is the same as the one normal to the single copy field $A_{\mu} = \Phi k_{\mu}$. In the
spirit of the double copy, these allow us to predict the existence of a trapped region using quantities
that can be defined using just the single and zeroth copy fields in flat space. Explicitly, the
term $\frac{\Phi}{2} \mathring{\theta}^{k}$ can be written as $1 / 2$ times the expansion along
$A_{\mu}$,
\begin{equation} \label{} 
    \begin{split}
        \frac{1}{2}\mathring{\theta}^{A} &= \frac{1}{2} \mathring{h}^{\mu \nu} \partial_{\mu}A_{\nu} \\
                                         &= \frac{1}{2} \mathring{h}^{\mu \nu} \partial_{\mu} \left(\Phi
                                         k_{\nu}\right) \\
                                         &= \frac{\Phi}{2} \mathring{h}^{\mu \nu} \partial_{\mu} k_{\nu},
    \end{split}
\end{equation}
since $\mathring{h}^{\mu \nu} k_{\nu} = 0$ by construction. In other words, since $A_{\mu}$ and $k_{\mu}$ are
parallel, evolving a co-dimension 2 volume element along either results in the same trajectory, albeit at differing
rates. The fractional rate of change is thus enhanced by a factor of $\Phi$ for evolution along $A_{\mu}$ when
compared to that along $k_{\mu}$.

To summarize, our findings here constitute a diagnostic test for the presence of a trapping horizon using flat
space computations, provided the existence of a classical double copy (in the Kerr-Schild sense) for fields
$A_{\mu}$ and $\Phi$. The test proceeds as follows: first pick a hypersurface $\mathcal{H}$ to test. Foliate it by
spacelike slices normal to $k_{\mu} = \frac{1}{\Phi} A_{\mu}$, and find the complementary null normal $l^{(0)}$ as
in \eqref{l0FromH}. Compute the expansions $\mathring{\theta}^{(k)}$ and $\mathring{\theta}^{(l^{(0)})}$. If the
combination $\mathring{\theta}^{(l^{(0)})} + \frac{\Phi}{2} \mathring{\theta}^{(k)}$ vanishes all along
$\mathcal{H}$, it is a trapping horizon.

\section{Examples} \label{sec:examples}
We now demonstrate how the approach outlined in the previous section applies to black holes whose horizon structure
as well as double copy description are already known. A large class of stationary black hole spacetimes whose metric can be put in the Kerr-Schild form
\eqref{kerrSchildMetric} is the Myers-Perry class. The double copy construction of the Myers-Perry metric in
general dimensions is given in \cite{Monteiro:2014cda}. We first demonstrate the double copy description of the
Schwarzschild horizon, which we hope illustrates our approach more transparently than the somewhat messier
expressions which arise in our analysis of the more general Myers-Perry metric.

\subsection{Schwarzschild black hole}
We write the Schwarzschild metric in Kerr-Schild coordinates as
\begin{equation} \label{SchMetricKerrSchild} 
    ds^{2} = -d\tau^{2} + dr^{2} + r^{2}d\theta^{2} + r^{2}\sin^{2}\theta d\phi^{2} + \frac{2M}{r} \left(-d\tau
    - dr\right)^{2}.
\end{equation}
The first four terms in \eqref{SchMetricKerrSchild} constitute the flat metric $\eta_{\mu \nu}$ in spherical polar
coordinates. The zeroth copy $\Phi$ and the Kerr-Schild vector $k_{\mu}$ are
\begin{equation} \label{} 
    \Phi = \frac{2M}{r}, \qquad k_{\mu} dx^{\mu} = -d\tau - dr,
\end{equation}
from which we can also construct the single copy gauge field
\begin{equation} \label{} 
    A_{\mu} = \Phi k_{\mu} = -\frac{2M}{r} \left(d\tau + dr\right).
\end{equation}
This is the gauge field of a static point charge, albeit in an unusual gauge.

The negative signs in $k_{\mu}$ ensure that its integral curves are future-directed and ingoing. Note that while
the spatial coordinates are precisely the same as the usual spherical polar coordinates, the time coordinate $\tau$
here relates to the usual Schwarzschild $t$ coordinate via
\begin{equation} \label{} 
    d\tau = dt + \frac{2M}{r} dr.
\end{equation}
The Kerr-Schild coordinate patch thus covers the same region as the infalling Eddington-Finkelstein coordinate
patch, which is obtained by one further coordinate transformation from $\tau$ to $u$ given by $du = d\tau + dr$.

\begin{figure}[h]
    \centering
    \begin{subfigure}[t]{0.60\textwidth}
        \centering
        \def\svgwidth{\columnwidth}
        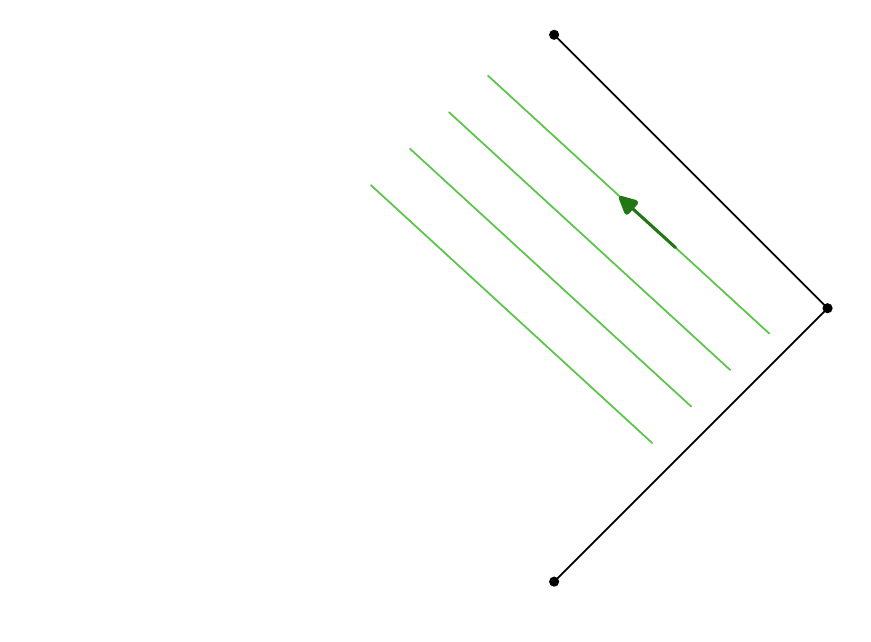
        \caption{Penrose diagram for the infalling coordinate patch of the Schwarzschild solution. Blue curves
        represent the family of test hypersurfaces $\mathcal{H}_{(r)}$ at constant $r$, green lines are the
    integral curves of $k^{\mu}$. $\mathcal{S}_{1}$ and $\mathcal{S}_{2}$ are leaves of the foliation of
$\mathcal{H}_{1}$ and $\mathcal{H}_{2}$ normal to $k^{\mu}$.}
    \end{subfigure}\hspace{2em}%
    \begin{subfigure}[t]{0.27\textwidth}
        \centering
        \def\svgwidth{\columnwidth}
        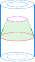
        \caption{Two of the constant $r$ hypersurfaces with one angular direction
        restored, showing the evolution from $\mathcal{S}_{1}$ to $\mathcal{S}_{2}$ along $k^{\mu}$.}
    \end{subfigure}
    \caption{Foliation of constant $r$ surfaces in Schwarzschild using $k^{\mu}$.} \label{figSchw}
\end{figure}

The event horizon $\mathcal{H}$ for the Schwarzschild black hole in Kerr-Schild coordinates is at $r = 2M$. We thus
look at the family of hypersurfaces $\mathcal{H}_{(r)}$ at constant $r$. The normals to these surfaces are along
the one-form $dr$. Normalizing the one-form as described in \eqref{HNorm}, we find that 
\begin{equation} \label{} 
    H_{\mu}dx^{\mu} = dr.
\end{equation}
On the auxiliary flat spacetime, the hypersurfaces $\mathcal{H}_{(r)}$ are foliated by co-dimension 2 spacelike
surfaces $\mathcal{S}_{\lambda}$ normal to $k_{\mu}$ and to $l^{(0)}_{\mu}$, which in Kerr-Schild coordinates is
\begin{equation} \label{} 
    \begin{split}
        l^{(0)}_{\mu} dx^{\mu}
        &= \frac{H_{\rho}H_{\sigma}\eta^{\rho\sigma}}{2} k_{\mu} dx^{\mu} + H_{\mu} dx^{\mu} \\
        &= \frac{\eta^{rr}}{2} \left(-d\tau - dr\right) + dr \\
        &= \frac{1}{2} \left(-d\tau + dr\right).
    \end{split}
\end{equation}
The null normal on Schwarzschild, from the general form \eqref{lFroml0Down}, is 
\begin{equation} 
    \begin{split}
        l_{\mu} dx^{\mu}
        &= l^{(0)}_{\mu} dx^{\mu} - \frac{\Phi}{2} k_{\mu} dx^{\mu} \\
        &= \frac{1}{2} \left(-d\tau  +dr\right) - \frac{M}{r} \left(-d\tau - dr\right) \\
        &= - \frac{1}{2} \left(1 - \frac{2M}{r}\right) d\tau + \frac{1}{2} \left(1 + \frac{2M}{r}\right) dr.
    \end{split}
\end{equation}
See figure \ref{figSchw} for an illustration of the hypersurfaces $\mathcal{H}_{(r)}$ and their foliation on the
Schwarzschild background.
In Schwarzschild spacetime, the vectors $l^{\mu}$ and $l^{(0)\mu}$ are parallel to geodesics with
respect to $g_{\mu \nu}$ and $\eta_{\mu \nu}$ respectively. In fact, $l^{\mu}$ lies along the generators of the
horizon at $r = 2M$.

Thus, in this case our prescription \eqref{lFroml0Down} for relating $l_{\mu}$ and $l_{\mu}^{(0)}$ maps a family of
null geodesics in flat spacetime to null geodesics on Schwarzschild. This map, however, is distinct from the one
proposed in \cite{Gonzo:2021drq}, which does not map freely-falling trajectories. Instead, it maps
on-shell trajectories of a test charge in the single copy gauge field to geodesics of a test particle in the double
copy metric.

Writing out the induced metric and computing expansions, we find on the flat spacetime
\begin{equation} \label{} 
    \mathring{\theta}^{(k)} = -\frac{2}{r}, \qquad \mathring{\theta}^{(l^{(0)})} = \frac{1}{r},
\end{equation}
and thus from our general result \eqref{thetaFromTheta0}
\begin{equation} \label{schwExpansionsFromFlatExpansions} 
    \theta^{(k)} = - \frac{2}{r}, \qquad \theta^{(l)} = \frac{1}{r} + \frac{M}{r} \left(- \frac{2}{r}\right)
    = \frac{r-2M}{r^{2}}.
\end{equation}
We see that, as expected on physical grounds, $\theta^{(k)}$ is always negative while $\theta^{(l)}$ is positive
outside the horizon at $r = 2M$ and negative inside. We can also explicitly check the outermost condition $\mathcal{L}_{k} \theta^{(l)} < 0$ on the marginal surface $r = 2M$,
\begin{equation}
    \begin{split}
        \mathcal{L}_{k} \theta^{(l)} \big\vert_{r = 2M}
        &= k^{\mu} \partial_{\mu} \theta^{(l)} \big\vert_{r = 2M} \\
        &= \left( \partial_{\tau} - \partial_{r} \right) \theta^{(l)} \big\vert_{r = 2M} \\
        &= - \frac{1}{4M^{2}}.
    \end{split}
\end{equation}
Thus, the surface $r = 2M$ is a future-outer trapping horizon \cite{Hayward:1993wb}.
From the perspective of the double copy construction we have
described, the change in sign of $\theta^{(l)}$ is a consequence of the interplay between
$\mathring{\theta}^{(l^{(0)})}, \mathring{\theta}^{(k)}$ and $\Phi$ as in \eqref{thetaFromTheta0}. As the horizon
$\Phi = 1$ is approached from the outside, the term $\frac{\Phi}{2} \mathring{\theta}^{(k)}$ starts to dominate over
$\mathring{\theta}^{(l^{(0)})}$.  

As noted in section \ref{sec:doubleCopyRelation}, the second term in our expression for $\theta^{(l)}$
\eqref{schwExpansionsFromFlatExpansions} can be equivalently written as $1 / 2$ the expansion of the single copy
field $A_{\mu}$. The expansion of the outgoing null vector $l^{\mu}$ with respect to $g_{\mu \nu}$ is thus related
to the expansion of $l^{(0) \mu}$ with respect to $\eta_{\mu \nu}$ via
\begin{equation} \label{} 
    \theta^{(l)} = \mathring{\theta}^{(l^{(0)})} + \frac{1}{2} \mathring{\theta}^{(A)}.
\end{equation}
The second term has a clear single copy interpretation, while the first term depends upon the arbitrary choice of
hypersurfaces $\mathcal{H}_{(r)}$ on the flat background $\eta_{\mu \nu}$.

We mentioned earlier in this section that $l^{\mu}$ is parallel to a family of geodesics. Explicitly, the vector
$\hat{l}^{\mu}$ defined as
\begin{equation} \label{} 
    \hat{l}^{\mu} = -\left( \frac{1}{2} - \frac{M}{r}\right)^{-1} l^{\mu},
\end{equation}
generates past-directed, radially-infalling null geodesics. The Schwarzschild metric can, in fact, be written in a
set of Kerr-Schild coordinates adapted to $\hat{l}_{\mu}$ instead of $k_{\mu}$, which cover the outgoing patch
instead (including the past horizon). This second set of Kerr-Schild coordinates map $g_{\mu \nu}$ to a distinct (albeit gauge-equivalent) single copy $\hat{A}_{\mu} = \Phi
\hat{l}_{\mu}$ living on a \emph{different} flat
background $\hat{\eta}_{\mu \nu}$. The same family of hypersurfaces $\mathcal{H}_{(r)}$ can now be used to locate
the past horizon of the eternal Schwarzschild black hole at $r = 2M$. The interested reader might find the
discussion of the various Kerr-Schild coordinate patches in \cite{Boyer:1966qh} useful. We omit delving into this
issue at present, and instead move on to the more general Myers-Perry class of black holes.

\subsection{Myers-Perry black holes}
The Myers-Perry spacetimes in arbitrary dimension $D \geq 4$ generalize the $D = 4$ Kerr black hole to arbitrary
dimension. Its double copy description first appeared in \cite{Monteiro:2014cda}, which we re-state here. Just as
for the Kerr black hole, Myers-Perry spacetimes have in general two horizons, an outer event horizon and an inner
Cauchy horizon. 

For $D \geq 4$, there are $(D-2) / 2$ independent rotation planes for even $D$, and
$(D-1)/2$ for odd $D$ (see for instance \cite{Myers:2011yc} or appendix E of \cite{Frolov:2017kze}). Call this number of independent rotations $m$. The black hole can be parametrized by
$m$ spin parameters $a_{i}$ and a mass $M$. In even dimensions we use coordinates $\left(\tau, x_{1}, y_{1}, \dots
x_{m}, y_{m}\right)$ where the pair $\left(x_{i}, y_{i}\right)$ spans the rotation plane corresponding to $a_{i}$.
In odd dimensions an additional spacelike coordinate $z$ is added. The relation between these Kerr-Schild
coordinates and the usual Myers-Perry coordinates $\left(t, r, \phi_{i}, \mu_{i}\right)$ can be found in, for
instance, appendix E.3 of \cite{Frolov:2017kze}.

The radial variable $r$ solves the equation
\begin{equation} \label{} 
    \left(1 - \varepsilon\right) \frac{z^{2}}{r^{2}} + \sum_{i = 1}^{m} \frac{x_{i}^{2} + y_{i}^{2}}{r^{2} +
    a_{i}^{2}} = 1,
\end{equation}
where $\varepsilon = 0$ for even $D$ and $\varepsilon = 1$ for odd $D$. Surfaces of constant $r$ within a constant
time-slice are thus ellipsoids, which become spheres in the non-rotating limit. We also define the functions $U$
and $V$,
\begin{equation} \label{} 
    V = \frac{1}{r^{1 + \varepsilon}} \prod_{i = 1}^{m}\left(r^{2} + a_{i}^{2}\right), \qquad U = V\left(1 -
    \sum_{i = 1}^{m} \frac{a_{i}^{2} \left(x_{i}^{2} + y_{i}^{2}\right)}{r^{2} + a_{i}^{2}}\right).
\end{equation}
With these definitions, the metric in Kerr-Schild coordinates is
\begin{multline} \label{} 
    ds^{2} = -d\tau^{2} + \left(1-\varepsilon \right) dz^{2} + \sum_{i = 1}^{m} \left(dx_{i}^{2} +
    dy_{i}^{2}\right) \\
    + \frac{2M}{U} \left( -d\tau - \left(1 - \varepsilon\right) \frac{z\, dz}{r} - \sum_{i = 1}^{m} \frac{r
    \left(x_{i}\, dx_{i} + y_{i}\, dy_{i}\right)}{r^{2} + a_{i}^{2}} - \sum_{i = 1}^{m} \frac{a_{i} \left(x_{i} \,
    dy_{i} - y_{i} \, dx_{i}\right)}{r^{2} + a_{i}^{2}}\right)^{2}.
\end{multline}
The first line constitutes the flat metric $\eta_{\mu \nu}$. The zeroth copy $\Phi$ and the Kerr-Schild vector
$k_{\mu}$ are
\begin{equation} \label{} 
    \Phi = \frac{2M}{U}, \qquad k_{\mu} dx^{\mu} = -d\tau - \left(1 - \varepsilon\right) \frac{z\,dz}{r} -
    \sum_{i = 1}^{m} \frac{r\left(x_{i}\, dx_{i} + y_{i} \, dy_{i}\right)}{r^{2} + a_{i}^{2}} - \sum_{i = 1}^{m}
    \frac{a_{i} \left(x_{i} \, dy_{i} - y_{i}\, dx_{i}\right)}{r^{2} + a_{i}^{2}}.
\end{equation}
The single copy field can be constructed using $A_{\mu} = \Phi k_{\mu}$.

Much like the choice of coordinates in Schwarzschild, the radial variable $r$ here is defined such that the
horizons are always a surface of constant $r$. Specifically, the horizons solve the equation $V - 2M = 0$. We thus
look at the family of hypersurfaces $\mathcal{H}_{(r)}$ at constant $r$. The normals to this family are along the
one-form $dr$, which in Kerr-Schild coordinates is
\begin{equation} \label{} 
    dr = \frac{V}{U} \left( \sum_{i=1}^{m} \frac{r\left(x_{i}\, dx_{i} + y_{i} \, dy_{i}\right)}{r^{2} +
    a_{i}^{2}} + \left(1-\varepsilon\right) \frac{z\,dz}{r}\right).
\end{equation}
Normalizing using \eqref{HNorm}, we find $H_{\mu}dx^{\mu} = dr$. 

The foliation of $\mathcal{H}_{(r)}$ in flat spacetime is given by the null normals $k_{\mu}$ and $l^{(0)}_{\mu}$,
\begin{equation} \label{} 
    \begin{split}
        l^{(0)}_{\mu} dx^{\mu} &= \frac{H_{\rho} H_{\sigma} \eta^{\rho \sigma}}{2} k_{\mu} dx^{\mu} + H_{\mu}
        dx^{\mu} \\
        &= \frac{V}{2U} k_{\mu} dx^{\mu} + dr. \\
    \end{split}
\end{equation}
Similar to the Schwarzschild case, we can find the corresponding null normal on the Myers-Perry background using
\eqref{lFroml0Down},
\begin{equation} \label{} 
    l_{\mu} dx^{\mu} = \frac{V - 2M}{2U} k_{\mu} dx^{\mu} + dr.
\end{equation}
We calculate the flat space expansions,
\begin{equation} \label{} 
    \begin{split}
        \mathring{\theta}^{(k)} &= \frac{\varepsilon}{r} - \frac{2}{r} \sum_{i = 1}^{m} \left[
        \frac{r^{2}}{r^{2} + a_{i}^{2}} + \frac{V}{U} \frac{a_{i}^{2} r^{2} \left(x_{i}^{2} +
    y_{i}^{2}\right)}{\left(r^{2} + a_{i}^{2}\right)^{3}}\right], \\
            \mathring{\theta}^{(l^{(0)})} &= - \frac{V}{2U} \mathring{\theta}^{(k)},
    \end{split}
\end{equation}
from which we obtain using \eqref{thetaFromTheta0},
\begin{equation} \label{} 
    \begin{split}
        \theta^{(k)} &= \frac{\varepsilon}{r} - \frac{2}{r} \sum_{i = 1}^{m} \left[
        \frac{r^{2}}{r^{2} + a_{i}^{2}} + \frac{V}{U} \frac{a_{i}^{2} r^{2} \left(x_{i}^{2} +
    y_{i}^{2}\right)}{\left(r^{2} + a_{i}^{2}\right)^{3}}\right], \\
            \theta^{(l)} &= -\frac{V - 2M}{2U} \theta^{(k)}.
    \end{split}
\end{equation}
Since $\mathring{\theta}^{(k)} < 0$, we see that $\theta^{(l)}$ vanishes whenever $V = 2M$.

Just as for Schwarzschild, the term $\frac{M}{U} \theta^{(k)}$ can be written as $\frac{1}{2}
\mathring{\theta}^{(A)}$, using the single-copy gauge field $A_{\mu} = \frac{2M}{U} k_{\mu}$. The past horizon can
also be similarly found, using Kerr-Schild coordinates adapted to $l_{\mu}$ instead of $k_{\mu}$.
See section 3.3 of \cite{Myers:1986un} for a discussion of the Kerr-Schild coordinate patch adapted to $l_{\mu}$.

\section{Discussion} \label{sec:discussion}
We have demonstrated a procedure for finding hoirzons within the Kerr-Schild classical double copy as first proposed in \cite{Monteiro:2014cda}.
The procedure is as follows: pick a hypersurface $\mathcal{H}$ to test, then foliate the surface using $k_\mu$.  Find the complementary null normal $l^{(0)}_\mu$ on the flat $\eta_{\mu\nu}$ background, and compute its expansion $\mathring{\theta}^{(l^{(0)})}$.  If the expansion $\mathring{\theta}^{(A)}$ of the single copy field $A_\mu=\Phi k_\mu$ equals $-2\mathring{\theta}^{(l^{(0)})}$ everywhere along the test hypersurface, then the double copy will have a trapping horizon there.

Importantly, our procedure relies on information available in the single copy alone.  The null normal $l^{(0)}_\mu$ is an object on the flat $\eta_{\mu\nu}$ background.  We do use the zeroth copy, in that $k_\mu=A_\mu/\Phi$, but no double-copy specific information is required.  We should point out that our perspective here assumes the single copy $A_\mu$ and the zeroth copy $\Phi$ live on the flat background $\eta_{\mu\nu}$.

Additionally, what we have actually found are marginally trapped surfaces. When we choose both $k_\mu$ and $l_\mu$ to be future directed, the future outer marginally trapped surface which occurs where $\theta^{(l)}=0$ will indicate an event horizon, but only if we additionally assume e.g. the null energy condition. As in the famous singularity theorems \cite{Penrose:1964wq}, future-directed light rays following a null congruence with zero expansion will inevitably collapse to a singularity due to the null Raychaudhuri equation. Allowing matter which disobeys the null energy condition could let the expansion turn positive again, letting the light escape.

Our horizon condition $\mathring{\theta}^{(l^{(0)})}=-\mathring{\theta}^{(A)}/2$ assumed the charge for the single copy gauge field was $M$.  If we wish to allow for the replacement $M\rightarrow q$, then the condition should be written as $\mathring{\theta}^{(l^{(0)})}=-M\mathring{\theta}^{(A)}/2q$.  Importantly, this refined condition indicates that changing the sign of $q$ in the single copy does not change the sign of the mass in the double copy; any choice of point charge in the $U(1)$ sector of the gauge theory results in a Schwarzschild black hole with a horizon.

Throughout this paper, our approach has been to consider the codimension-1 surface $\mathcal{H}$ as being at the same coordinate location in both the flat spacetime of the single copy and the curved spacetime of the double copy.  The paths tangent to $k^\mu$ are the same in both spacetimes, while the $l^\mu$ and $l^{(0)\mu}$ paths are chosen by the same leaves of $k^\mu$ in the hypersurface $\mathcal{H}$. Even though $l^\mu$ for the cases we examined ends up being tangent to a set of geodesics, we have not followed previous work on the double copy of geodesics \cite{Gonzo:2021drq}.  In \cite{Gonzo:2021drq}, the authors instead replace the conserved charges associated with the curved-space geodesics with conserved charges for particles moving in the single-copy gauge field.

Our work does extend to non-vacuum Kerr-Schild spacetimes; as an example, it is straightforward to show that metrics of the general form $ds^2=g_{tt}(r) dt^2-1/g_{tt}(r)\, dr^2+r^2d\Omega^2$, as studied in \cite{Easson:2020esh}, have a Kerr-Schild form and thus a simple single-copy.  Following our procedure here, the trapping horizons unsurprisingly appear where $\theta^{(l)} = g_{tt}/r=0$.  Just as in the current work, the data needed to find this result is entirely in the single copy and its flat background.

For the vacuum Schwarzschild, Kerr, and Myers-Perry black holes we investigated, the $k^\mu$ in the Kerr-Schild coordinates is also a (doubled) principal null vector of the Weyl tensor.  These spacetimes are Petrov type D, so their Weyl tensor should have two doubled principal null vectors. When we have correctly chosen the $\mathcal{H}$ surface to be along the horizon, then $l^\mu$ actually turns out to be exactly this second doubled principal null vector.  So far we have investigated horizons only for single Kerr-Schild spacetimes, but the classical double copy extends to multi-Kerr-Schild \cite{Chawla:2022ogv, Luna:2015paa} as well as Weyl formalisms \cite{Luna:2018dpt,Godazgar:2020zbv}. We defer further investigations of the relationship between local horizon definitions and these more general classical double copy formalisms to future work.

\acknowledgments
The authors thank Kwinten Fransen, Damien Easson, Tucker Manton, and Raphaela Wutte for helpful discussions.
The authors are supported by the U.S. Department of Energy under grant number DE-SC0019470 and by the
Heising-Simons Foundation ``Observational Signatures of Quantum Gravity'' collaboration grant 2021-2818.

\begin{appendices}
\addtocontents{toc}{\protect\setcounter{tocdepth}{1}}
\section{Calculation of curved spacetime expansions from flat spacetime expansions} \label{app:ExpansionCalculation} 
In this appendix we give a detailed derivation of our main result relating the expansions on the $\eta_{\mu \nu}$ and the $g_{\mu \nu}$ spacetimes \eqref{thetaFromTheta0} for a generic Kerr-Schild metric
\eqref{kerrSchildMetric}
\begin{equation}
    g_{\mu \nu} = \eta_{\mu \nu} + \Phi k_{\mu} k_{\nu}.
\end{equation}
We will use coordinate-basis methods throughout. Computing expansions is also possible in an orthonormal basis using the Newman-Penrose formalism, where the expansion appears as one of the various spin coefficients. Our main result \eqref{thetaFromTheta0} is corroborated by just such a calculation of ``shifted spin coefficients'' performed by \cite{BenTov:2017kyf} in a different context.

First, for notational simplicity, we define the vectors $l^{(0)\mu}$ and $l^{\mu}$, which live respectively on the $\eta_{\mu \nu}$ and $g_{\mu \nu}$ backgrounds,
\begin{align}
    l^{(0) \mu} &\coloneqq \eta^{\mu \nu} l^{(0)}_{\nu} \\
    l^{\mu} &\coloneqq g^{\mu \nu} l_{\mu}.
\end{align}
Thus, including \eqref{lFroml0Down}, we have the relations
\begin{align}
    l^{\mu} &= l^{(0) \mu} + \frac{\Phi}{2} k^{\mu} \label{lFroml0Up} \\
    l_{\mu} &= l^{(0)}_{\mu} - \frac{\Phi}{2} k_{\mu}.
\end{align}
In terms of $l^{\mu}$ and $l^{(0) \mu}$, we can write the induced (inverse) metrics $h^{\mu \nu}$ \eqref{flatInducedMetric} and $\mathring{h}^{\mu \nu}$ \eqref{curvedInducedMetric} more succinctly as
\begin{align}
    h^{\mu \nu} &= g^{\mu \nu} + k^\mu l^\nu + l^\mu k^\nu \\
    \mathring{h}^{\mu \nu} &= \eta^{\mu \nu} + k^\mu l^{(0) \nu} + l^{(0) \mu} k^\nu.
\end{align}
Using the explicit inverse Kerr-Schild metric \eqref{kerrSchildMetricInverse} and the relation \eqref{lFroml0Up}, 
\begin{equation}
    \begin{split}
        h^{\mu \nu}
        &= \eta^{\mu \nu} - \Phi k^\mu k^\nu + k^\mu \left( l^{(0) \nu} + \frac{\Phi}{2} k^\nu \right) + \left( l^{(0) \mu} + \frac{\Phi}{2} k^\mu \right) k^\nu \\
        &= \eta^{\mu \nu} + k^\mu l^{(0) \nu} + l^{(0) \mu} k^\nu \\
        &= \mathring{h}^{\mu \nu}.
    \end{split}
\end{equation}
We have thus found that the metric on $\mathcal{S}_\lambda$ induced from the flat and the curved background are equal. We henceforth drop the circular accent on $\mathring{h}^{\mu \nu}$.

To relate the flat- and curved-spacetime expansions defined in \eqref{flatExpansions} and \eqref{curvedExpansions}, we will also need to relate the derivatives $\mathring{\nabla}_\mu$ and $\nabla_\mu$. It is a standard result in differential geometry that the difference between the action of any two covariant derivatives on a one-form is a linear map (see e.g. chapter 3 of \cite{Wald:1984rg}), 
\begin{equation}
    \nabla_\mu V_\rho - \mathring{\nabla}_\mu V_\rho = -\Gamma_{\mu \rho}^{\sigma} V_{\sigma}.
\end{equation}
We can thus write
\begin{align}
    \begin{split} \label{thetaKIntermediate}
        \theta^{(k)}
        &= h^{\mu \nu} \nabla_\mu k_\nu \\
        &= h^{\mu \nu} \mathring{\nabla}_\mu k_\mu - h^{\mu \nu} \Gamma_{\mu \nu}^\sigma k_\sigma \\
        &= \mathring{\theta}^{(k)} - h^{\mu \nu} \Gamma_{\mu \nu}^\sigma k_\sigma,
    \end{split} \\
    \intertext{and}
    \begin{split} \label{thetaLIntermediate}
        \theta^{(l)}
        &= h^{\mu \nu} \nabla_\mu l_\nu \\
        &= h^{\mu \nu} \mathring{\nabla}_\mu l_\nu - h^{\mu \nu} \Gamma_{\mu \nu}^\sigma l_\sigma \\
        &= h^{\mu \nu} \mathring{\nabla}_\mu \left(l_\nu^{(0)} - \frac{\Phi}{2} k_\nu \right) - h^{\mu \nu} \Gamma_{\mu \nu}^\sigma l_\sigma \\
        &= \mathring{\theta}^{(l^{(0)})} -  h^{\mu \nu} \Gamma_{\mu \nu}^\sigma l_\sigma - h^{\mu \nu} \mathring{\nabla}_\mu \left( \frac{\Phi}{2} k_\sigma \right) \\
        &= \mathring{\theta}^{(l^{(0)})} -  h^{\mu \nu} \Gamma_{\mu \nu}^\sigma l_\sigma - \frac{\Phi}{2}\mathring{\theta}^{(k)}.
    \end{split}
\end{align}
We demonstrated that the last equality holds in section \ref{sec:doubleCopyRelation}. Clearly, we will need the contractions $h^{\mu \nu} \Gamma_{\sigma \mu \nu} k^\sigma$ and $h^{\mu \nu} \Gamma_{\sigma \mu \nu} l^\sigma$.

First, from the metric compatibility of the two derivative operators with $g_{\mu \nu}$ and $\eta_{\mu \nu}$ respectively, we can write
\begin{equation}
    \Gamma_{\sigma \mu \nu} = g_{\sigma \lambda} \Gamma^\lambda_{\mu \nu} = \frac{1}{2}\left( \mathring{\nabla}_{\nu} \left[\Phi k_{\mu} k_{\sigma}\right] +
    \mathring{\nabla}_{\mu} \left[\Phi k_{\nu} k_{\sigma}\right] - \mathring{\nabla}_{\sigma}\left[\Phi k_{\mu} k_{\nu}\right]\right).
\end{equation}
Let us simplify $h^{\mu \nu}\Gamma_{\sigma \mu \nu}$,
\begin{equation} \label{hGamma}
    h^{\mu \nu} \Gamma_{\sigma \mu \nu} = \eta^{\mu \nu} \Gamma_{\sigma \mu \nu} + 2 k^\mu l^{(0)\nu} \Gamma_{\sigma \mu \nu},
\end{equation}
where we used the index symmetry of $\Gamma_{\sigma \mu \nu}$. Using the metric compatibility of $\mathring{\nabla}$ with $\eta_{\mu \nu}$ and the nullness of $k_\mu$, the first term in \eqref{hGamma} simplifies as
\begin{equation}
    \frac{1}{2} \eta^{\mu \nu}\left( \mathring{\nabla}_{\nu} \left[\Phi k_{\mu} k_{\sigma}\right] +
\mathring{\nabla}_{\mu} \left[\Phi k_{\nu} k_{\sigma}\right] - \mathring{\nabla}_{\sigma}\left[\Phi k_{\mu} k_{\nu}\right]\right) = \mathring{\nabla}_\mu \left[ \Phi k^\mu k_\sigma \right].
\end{equation}
To simplify the second term of \eqref{hGamma}, we compute
\begin{equation} \label{} 
    \begin{split}
        k^{\mu} \Gamma_{\sigma \mu \nu} &= \frac{1}{2} k^{\mu} \mathring{\nabla}_{\mu} \left[\Phi k_{\nu} k_{\sigma}\right] + \frac{1}{2} k^{\mu}
        \mathring{\nabla}_{\nu} \left[\Phi k_{\mu} k_{\sigma}\right] - \frac{1}{2} k^{\mu} \mathring{\nabla}_{\sigma} \left[\Phi k_{\mu} k_{\nu}\right] 
        \\
                           &= \frac{1}{2} \Phi k_{\nu} k^{\mu} \mathring{\nabla}_{\mu}k_{\sigma} + \frac{1}{2} \Phi k_{\sigma} k^{\mu}
                           \mathring{\nabla}_{\mu} k_{\nu} + \frac{1}{2} k_{\nu} k_{\sigma} k^{\mu} \mathring{\nabla}_{\mu} \Phi + \frac{1}{2}
                           k^{\mu}k_{\mu} \mathring{\nabla}_{\nu}\left[\Phi k_{\sigma}\right] \\
                           &\phantom{=\ }\ + \frac{1}{2} \Phi k_{\sigma} k^{\mu}
                           \mathring{\nabla}_{\nu} k_{\mu} - \frac{1}{2} k^{\mu} k_{\mu} \mathring{\nabla}_{\sigma} \left[\Phi k_{\nu}\right] -
                           \frac{1}{2} \Phi k_{\nu} k^{\mu} \mathring{\nabla}_{\sigma} k_{\mu} \\
                           &= \frac{1}{2} k_{\nu} k_{\sigma} k^{\mu} \mathring{\nabla}_{\mu}\Phi.
    \end{split}
\end{equation}
The last line here follows from the previous by observing that the nullness and geodeticity of $k_\mu$ forces all
but the third term to vanish. The second term of \eqref{hGamma} is therefore
\begin{equation}
    2 k^\mu l^{(0) \nu} \Gamma_{\sigma \mu \nu} = l^{(0) \nu}  k_\nu k_\sigma k^\mu \mathring{\nabla}_\mu \Phi = - k_\sigma k^\mu \mathring{\nabla}_\mu \Phi,
\end{equation}
where we used the normalization condition $l^{(0)\mu} k_\mu = -1$. We have thus simplified \eqref{hGamma} to
\begin{equation}
    \begin{split}
        h^{\mu \nu} \Gamma_{\sigma \mu \nu}
        &= \mathring{\nabla}_\mu \left[ \Phi k^\mu k_\sigma \right] - k_\sigma k^\mu \mathring{\nabla}_\mu \Phi \\
        &= \Phi \mathring{\nabla}_\mu \left[ k^\mu k_\sigma \right] \\
        &= \Phi k_\sigma \mathring{\nabla}_\mu k^\mu,
    \end{split}
\end{equation}
where to obtain the final equality we used the geodeticity of $k_\sigma$. In just a moment we will need also the equality $\mathring{\theta}^{(k)} = \mathring{\nabla}_\mu k^\mu$, which we now show. The difference between the two terms can be simplified as,
\begin{equation}
    \begin{split}
        \mathring{\theta}^{(k)} - \mathring{\nabla}_\mu k^\mu
        &= \left( h^{\mu \nu} - \eta^{\mu \nu} \right) \mathring{\nabla}_\mu k_\nu \\
        &= \left(k^\mu l^{(0)\nu} + l^{(0)\mu} k^\nu \right) \mathring{\nabla}_\mu k_\nu \\
        &= l^{(0) \nu} k^\mu \mathring{\nabla}_\mu k_\nu + \frac{1}{2} l^{(0) \mu} \mathring{\nabla}_\mu \left(k^\nu k_\nu \right) \\
        &= 0,
    \end{split}
\end{equation}
where the first term vanishes due to geodeticity and the second due to the nullness of $k^\mu$. We can finally compute the two contractions we needed,
\begin{align}
    h^{\mu \nu} \Gamma_{\sigma \mu \nu} k^\sigma &= \Phi \left(k^\sigma k_\sigma\right) \mathring{\nabla}_\mu k^\mu = 0, \label{hGammaK} \\
    h^{\mu \nu} \Gamma_{\sigma \mu \nu} l^\sigma &= \Phi \left(l^\sigma k_\sigma\right) \mathring{\nabla}_\mu k^\mu = -\Phi \mathring{\theta}^{(k)}. \label{hGammaL}
\end{align}
Substituting \eqref{hGammaK} into \eqref{thetaKIntermediate} and \eqref{hGammaL} into \eqref{thetaLIntermediate},
\begin{align}
    \theta^{(k)} &= \mathring{\theta}^{(k)} - h^{\mu \nu}\Gamma_{\sigma \mu \nu} k^\sigma = \mathring{\theta}^{(k)}. \\
    \begin{split}
        \theta^{(l)} &= \mathring{\theta}^{(l^{(0)})} -  h^{\mu \nu} \Gamma_{\mu \nu}^\sigma l_\sigma - \frac{\Phi}{2}\mathring{\theta}^{(k)} = \mathring{\theta}^{(l^{(0)})} + \Phi\mathring{\theta}^{(k)} - \frac{\Phi}{2}\mathring{\theta}^{(k)} \\
        &= \mathring{\theta}^{(l^{(0)})} + \frac{\Phi}{2}\mathring{\theta}^{(k)}.
    \end{split}
\end{align}

\end{appendices}

\bibliographystyle{jhep.bst}
\bibliography{5DDoubleCopy,LocalHorizons} 

\end{document}

%% file: GenericPenrose.pdf_tex
\begingroup%
  \makeatletter%
  \providecommand\color[2][]{%
    \errmessage{(Inkscape) Color is used for the text in Inkscape, but the package 'color.sty' is not loaded}%
    \renewcommand\color[2][]{}%
  }%
  \providecommand\transparent[1]{%
    \errmessage{(Inkscape) Transparency is used (non-zero) for the text in Inkscape, but the package 'transparent.sty' is not loaded}%
    \renewcommand\transparent[1]{}%
  }%
  \providecommand\rotatebox[2]{#2}%
  \newcommand*\fsize{\dimexpr\f@size pt\relax}%
  \newcommand*\lineheight[1]{\fontsize{\fsize}{#1\fsize}\selectfont}%
  \ifx\svgwidth\undefined%
    \setlength{\unitlength}{314.52918826bp}%
    \ifx\svgscale\undefined%
      \relax%
    \else%
      \setlength{\unitlength}{\unitlength * \real{\svgscale}}%
    \fi%
  \else%
    \setlength{\unitlength}{\svgwidth}%
  \fi%
  \global\let\svgwidth\undefined%
  \global\let\svgscale\undefined%
  \makeatother%
  \begin{picture}(1,0.90335506)%
    \lineheight{1}%
    \setlength\tabcolsep{0pt}%
    \put(0,0){\includegraphics[width=\unitlength,page=1]{GenericPenrose.pdf}}%
    \put(0.56267858,0.55064846){\color[rgb]{0,0,0}\makebox(0,0)[lt]{\lineheight{1.25}\smash{\begin{tabular}[t]{l}\textcolor[HTML]{227614}{$k^\mu$}\end{tabular}}}}%
    \put(0.70550568,0.71927658){\color[rgb]{0,0,0}\makebox(0,0)[lt]{\lineheight{1.25}\smash{\begin{tabular}[t]{l}$\mathscr{I}^+$\end{tabular}}}}%
    \put(0.70458417,0.21431379){\color[rgb]{0,0,0}\makebox(0,0)[lt]{\lineheight{1.25}\smash{\begin{tabular}[t]{l}$\mathscr{I}^-$\end{tabular}}}}%
    \put(0.2420088,0.43270082){\color[rgb]{0,0,0}\makebox(0,0)[lt]{\lineheight{1.25}\smash{\begin{tabular}[t]{l}\textcolor[HTML]{2b8ce2}{$\mathcal{H}_{(\xi)}$}\end{tabular}}}}%
    \put(0.51199803,0.89343343){\color[rgb]{0,0,0}\makebox(0,0)[lt]{\lineheight{1.25}\smash{\begin{tabular}[t]{l}$i^+$\end{tabular}}}}%
    \put(0.51660535,0.00237687){\color[rgb]{0,0,0}\makebox(0,0)[lt]{\lineheight{1.25}\smash{\begin{tabular}[t]{l}$i^-$\end{tabular}}}}%
    \put(0.96259433,0.45942342){\color[rgb]{0,0,0}\makebox(0,0)[lt]{\lineheight{1.25}\smash{\begin{tabular}[t]{l}$i^0$\end{tabular}}}}%
    \put(0,0){\includegraphics[width=\unitlength,page=2]{GenericPenrose.pdf}}%
    \put(0.42676264,0.70637607){\color[rgb]{0.91764706,0.08235294,0.17254902}\makebox(0,0)[lt]{\lineheight{1.25}\smash{\begin{tabular}[t]{l}$\mathcal{S}_1$\end{tabular}}}}%
    \put(0.33598735,0.7649251){\color[rgb]{0.99215686,0.05490196,1}\makebox(0,0)[lt]{\lineheight{1.25}\smash{\begin{tabular}[t]{l}$\mathcal{S}_2$\end{tabular}}}}%
  \end{picture}%
\endgroup%

%% file: GenericCylinderFoliation.pdf_tex
\begingroup%
  \makeatletter%
  \providecommand\color[2][]{%
    \errmessage{(Inkscape) Color is used for the text in Inkscape, but the package 'color.sty' is not loaded}%
    \renewcommand\color[2][]{}%
  }%
  \providecommand\transparent[1]{%
    \errmessage{(Inkscape) Transparency is used (non-zero) for the text in Inkscape, but the package 'transparent.sty' is not loaded}%
    \renewcommand\transparent[1]{}%
  }%
  \providecommand\rotatebox[2]{#2}%
  \newcommand*\fsize{\dimexpr\f@size pt\relax}%
  \newcommand*\lineheight[1]{\fontsize{\fsize}{#1\fsize}\selectfont}%
  \ifx\svgwidth\undefined%
    \setlength{\unitlength}{18.59033239bp}%
    \ifx\svgscale\undefined%
      \relax%
    \else%
      \setlength{\unitlength}{\unitlength * \real{\svgscale}}%
    \fi%
  \else%
    \setlength{\unitlength}{\svgwidth}%
  \fi%
  \global\let\svgwidth\undefined%
  \global\let\svgscale\undefined%
  \makeatother%
  \begin{picture}(1,1.59145711)%
    \lineheight{1}%
    \setlength\tabcolsep{0pt}%
    \put(0,0){\includegraphics[width=\unitlength,page=1]{GenericCylinderFoliation.pdf}}%
    \put(0.86372523,0.91399655){\color[rgb]{0.13333333,0.4627451,0.07843137}\makebox(0,0)[lt]{\lineheight{1.25}\smash{\begin{tabular}[t]{l}$k^\mu$\end{tabular}}}}%
    \put(0.05731065,0.91399655){\color[rgb]{0.13333333,0.4627451,0.07843137}\makebox(0,0)[lt]{\lineheight{1.25}\smash{\begin{tabular}[t]{l}$k^\mu$\end{tabular}}}}%
    \put(0,0){\includegraphics[width=\unitlength,page=2]{GenericCylinderFoliation.pdf}}%
    \put(0.43588508,1.24121476){\color[rgb]{0.99215686,0.05490196,1}\makebox(0,0)[lt]{\lineheight{1.25}\smash{\begin{tabular}[t]{l}$\mathcal{S}_2$\end{tabular}}}}%
    \put(0.43588508,0.44736294){\color[rgb]{0.91764706,0.08235294,0.17254902}\makebox(0,0)[lt]{\lineheight{1.25}\smash{\begin{tabular}[t]{l}$\mathcal{S}_1$\end{tabular}}}}%
  \end{picture}%
\endgroup%

%% file: SchwPenrose.pdf_tex
\begingroup%
  \makeatletter%
  \providecommand\color[2][]{%
    \errmessage{(Inkscape) Color is used for the text in Inkscape, but the package 'color.sty' is not loaded}%
    \renewcommand\color[2][]{}%
  }%
  \providecommand\transparent[1]{%
    \errmessage{(Inkscape) Transparency is used (non-zero) for the text in Inkscape, but the package 'transparent.sty' is not loaded}%
    \renewcommand\transparent[1]{}%
  }%
  \providecommand\rotatebox[2]{#2}%
  \newcommand*\fsize{\dimexpr\f@size pt\relax}%
  \newcommand*\lineheight[1]{\fontsize{\fsize}{#1\fsize}\selectfont}%
  \ifx\svgwidth\undefined%
    \setlength{\unitlength}{417.66424942bp}%
    \ifx\svgscale\undefined%
      \relax%
    \else%
      \setlength{\unitlength}{\unitlength * \real{\svgscale}}%
    \fi%
  \else%
    \setlength{\unitlength}{\svgwidth}%
  \fi%
  \global\let\svgwidth\undefined%
  \global\let\svgscale\undefined%
  \makeatother%
  \begin{picture}(1,0.73738444)%
    \lineheight{1}%
    \setlength\tabcolsep{0pt}%
    \put(0,0){\includegraphics[width=\unitlength,page=1]{SchwPenrose.pdf}}%
    \put(0.78349489,0.3831782){\color[rgb]{0.13333333,0.4627451,0.07843137}\makebox(0,0)[lt]{\lineheight{1.25}\smash{\begin{tabular}[t]{l}$k^\mu$\end{tabular}}}}%
    \put(0.78948531,0.56274829){\color[rgb]{0,0,0}\makebox(0,0)[lt]{\lineheight{1.25}\smash{\begin{tabular}[t]{l}$\mathscr{I}^+$\end{tabular}}}}%
    \put(0.78948531,0.18272634){\color[rgb]{0,0,0}\makebox(0,0)[lt]{\lineheight{1.25}\smash{\begin{tabular}[t]{l}$\mathscr{I}^-$\end{tabular}}}}%
    \put(0.38450357,0.3831782){\color[rgb]{0.08627451,0.2745098,0.44313725}\makebox(0,0)[lt]{\lineheight{1.25}\smash{\begin{tabular}[t]{l}$\mathcal{H}_{(r)}$\end{tabular}}}}%
    \put(0.62787224,0.00178994){\color[rgb]{0,0,0}\makebox(0,0)[lt]{\lineheight{1.25}\smash{\begin{tabular}[t]{l}$i^-$\end{tabular}}}}%
    \put(0.97183106,0.36929972){\color[rgb]{0,0,0}\makebox(0,0)[lt]{\lineheight{1.25}\smash{\begin{tabular}[t]{l}$i^0$\end{tabular}}}}%
    \put(0,0){\includegraphics[width=\unitlength,page=2]{SchwPenrose.pdf}}%
    \put(0.62787224,0.72436138){\color[rgb]{0,0,0}\makebox(0,0)[lt]{\lineheight{1.25}\smash{\begin{tabular}[t]{l}$i^+$\end{tabular}}}}%
    \put(0,0){\includegraphics[width=\unitlength,page=3]{SchwPenrose.pdf}}%
    \put(-0.00062302,0.72436138){\color[rgb]{0,0,0}\makebox(0,0)[lt]{\lineheight{1.25}\smash{\begin{tabular}[t]{l}$i^+$\end{tabular}}}}%
    \put(0.2657863,0.72991279){\color[rgb]{0,0,0}\makebox(0,0)[lt]{\lineheight{1.25}\smash{\begin{tabular}[t]{l}$r = 0$\end{tabular}}}}%
    \put(0,0){\includegraphics[width=\unitlength,page=4]{SchwPenrose.pdf}}%
    \put(0.11082356,0.53053881){\color[rgb]{0.06666667,0,0.07058824}\rotatebox{-45}{\makebox(0,0)[lt]{\lineheight{1.25}\smash{\begin{tabular}[t]{l}$r = 2M$\end{tabular}}}}}%
    \put(0.55604425,0.42807068){\color[rgb]{0.91764706,0.08235294,0.17254902}\makebox(0,0)[lt]{\lineheight{1.25}\smash{\begin{tabular}[t]{l}$\mathcal{S}_1$\end{tabular}}}}%
    \put(0.45728074,0.46398484){\color[rgb]{0.99215686,0.05490196,1}\makebox(0,0)[lt]{\lineheight{1.25}\smash{\begin{tabular}[t]{l}$\mathcal{S}_2$\end{tabular}}}}%
    \put(0.33158162,0.44602779){\color[rgb]{0.06666667,0,0.07058824}\rotatebox{45}{\makebox(0,0)[lt]{\lineheight{1.25}\smash{\begin{tabular}[t]{l}$r = 2M$\end{tabular}}}}}%
    \put(0.36222166,0.27914071){\color[rgb]{0.06666667,0,0.07058824}\rotatebox{-45}{\makebox(0,0)[lt]{\lineheight{1.25}\smash{\begin{tabular}[t]{l}$r = 2M$\end{tabular}}}}}%
  \end{picture}%
\endgroup%

%% file: SchwCylinderFoliation.pdf_tex
\begingroup%
  \makeatletter%
  \providecommand\color[2][]{%
    \errmessage{(Inkscape) Color is used for the text in Inkscape, but the package 'color.sty' is not loaded}%
    \renewcommand\color[2][]{}%
  }%
  \providecommand\transparent[1]{%
    \errmessage{(Inkscape) Transparency is used (non-zero) for the text in Inkscape, but the package 'transparent.sty' is not loaded}%
    \renewcommand\transparent[1]{}%
  }%
  \providecommand\rotatebox[2]{#2}%
  \newcommand*\fsize{\dimexpr\f@size pt\relax}%
  \newcommand*\lineheight[1]{\fontsize{\fsize}{#1\fsize}\selectfont}%
  \ifx\svgwidth\undefined%
    \setlength{\unitlength}{16.66454315bp}%
    \ifx\svgscale\undefined%
      \relax%
    \else%
      \setlength{\unitlength}{\unitlength * \real{\svgscale}}%
    \fi%
  \else%
    \setlength{\unitlength}{\svgwidth}%
  \fi%
  \global\let\svgwidth\undefined%
  \global\let\svgscale\undefined%
  \makeatother%
  \begin{picture}(1,1.76317591)%
    \lineheight{1}%
    \setlength\tabcolsep{0pt}%
    \put(0,0){\includegraphics[width=\unitlength,page=1]{SchwCylinderFoliation.pdf}}%
    \put(0.76838415,0.858903){\color[rgb]{0.13333333,0.4627451,0.07843137}\makebox(0,0)[lt]{\lineheight{1.25}\smash{\begin{tabular}[t]{l}$k^\mu$\end{tabular}}}}%
    \put(0.11939096,0.858903){\color[rgb]{0.13333333,0.4627451,0.07843137}\makebox(0,0)[lt]{\lineheight{1.25}\smash{\begin{tabular}[t]{l}$k^\mu$\end{tabular}}}}%
    \put(0.4066703,1.33780875){\color[rgb]{0.99215686,0.05490196,1}\makebox(0,0)[lt]{\lineheight{1.25}\smash{\begin{tabular}[t]{l}$\mathcal{S}_2$\end{tabular}}}}%
    \put(0,0){\includegraphics[width=\unitlength,page=2]{SchwCylinderFoliation.pdf}}%
    \put(0.40592409,0.42919285){\color[rgb]{0.91764706,0.08235294,0.17254902}\makebox(0,0)[lt]{\lineheight{1.25}\smash{\begin{tabular}[t]{l}$\mathcal{S}_1$\end{tabular}}}}%
  \end{picture}%
\endgroup%